\documentclass[aps,prl,twocolumn,superscriptaddress]{revtex4-1}

\usepackage{graphicx,graphics,epsfig}   % Include figure files
\usepackage{xcolor}
\usepackage{dcolumn}                    % Align table columns on decimal point
\usepackage{bm}                         % bold math
\usepackage{amsmath,amsfonts,amssymb}   % need for subequations
\usepackage{subfigure}                  % use for side-by-side figures
\usepackage{times}                      % obsolete??
\usepackage{bbm}
\usepackage[normalem]{ulem}

\usepackage{hyperref}
\hypersetup{
    colorlinks=true,
    linkcolor=blue,
    filecolor=blue,      
    urlcolor=blue,
    citecolor=blue,
}

\newcommand{\ket}[1]{\mbox{$ | #1 \rangle $}}

\newcommand{\jdam}[1]{\textcolor{blue}{#1}}

\newcommand{\kb}[2]{| #1 \rangle\langle #2 |}

\newcommand{\tr}[1]{\mathrm{tr} \left(#1\right)}

\newcommand{\s}{\sigma}

\newcommand{\id}{\mathbbm{1}}
\newcommand{\set}[1]{\{ #1 \}}

\begin{document}

\title{Tracing Information Flow from Open Quantum Systems}
%Simulation of Open Quantum Systems in Integrated Waveguides: Information Flow in Non-Markovian System
%Integrated Waveguides for the analysis of information flow in open systems
%Tracing Information in Open Systems/ in non-Markovian Systems
%Tracing Information Flow from Open Systems
%Investigating Information Flow in Open Systems
%Tracing Information Flow from Open (Quantum) systems in Integrated Waveguides

\author{Jan Dziewior}
	\affiliation{Max-Planck-Institut f\"{u}r Quantenoptik, Hans-Kopfermann-Stra{\ss}e 1, 85748 Garching, Germany}
	\affiliation{Department f\"{u}r Physik, Ludwig-Maximilians-Universit\"{a}t, Schellingstra{\ss}e 4, 80799 M\"{u}nchen, Germany}
	\affiliation{Munich Center for Quantum Science and Technology (MCQST), Schellingstra{\ss}e 4, 80799 M\"{u}nchen, Germany}
	
\author{Leonardo Ruscio}
	\affiliation{Max-Planck-Institut f\"{u}r Quantenoptik, Hans-Kopfermann-Stra{\ss}e 1, 85748 Garching, Germany}
	\affiliation{Department f\"{u}r Physik, Ludwig-Maximilians-Universit\"{a}t, Schellingstra{\ss}e 4, 80799 M\"{u}nchen, Germany}
	\affiliation{Munich Center for Quantum Science and Technology (MCQST), Schellingstra{\ss}e 4, 80799 M\"{u}nchen, Germany}
	
\author{Lukas Knips}
	\affiliation{Max-Planck-Institut f\"{u}r Quantenoptik, Hans-Kopfermann-Stra{\ss}e 1, 85748 Garching, Germany}
	\affiliation{Department f\"{u}r Physik, Ludwig-Maximilians-Universit\"{a}t, Schellingstra{\ss}e 4, 80799 M\"{u}nchen, Germany}
	\affiliation{Munich Center for Quantum Science and Technology (MCQST), Schellingstra{\ss}e 4, 80799 M\"{u}nchen, Germany}
	
\author{Eric Meyer}
	\affiliation{Institut für Physik, Universität Rostock, Albert-Einstein-Str. 23, 18059 Rostock, Germany}
	
\author{Alexander Szameit}
	\affiliation{Institut für Physik, Universität Rostock, Albert-Einstein-Str. 23, 18059 Rostock, Germany}
	
\author{Jasmin D. A. Meinecke}
	\affiliation{Max-Planck-Institut f\"{u}r Quantenoptik, Hans-Kopfermann-Stra{\ss}e 1, 85748 Garching, Germany}
	\affiliation{Department f\"{u}r Physik, Ludwig-Maximilians-Universit\"{a}t, Schellingstra{\ss}e 4, 80799 M\"{u}nchen, Germany}
	\affiliation{Munich Center for Quantum Science and Technology (MCQST), Schellingstra{\ss}e 4, 80799 M\"{u}nchen, Germany}
	
\begin{abstract}
Open quantum systems are highly relevant, both for practical applications as well as for fundamental questions about the nature of information and its transfer, encompassing for example decoherence and memory effects.
Quantum mechanics introduces additional complexity to the transfer of information, e.g., storage of information in non-classical correlations.
Yet, some of these aspects tend to be neglected by the usual framework of open system research.
In this work we use photons in a waveguide array to implement a quantum simulation of the coupling of a qubit with a low-dimensional discrete environment.
Using the trace distance between quantum states as a measure of information, we analyze different types of information transfer. Extending the usual perspective which is focused on the system alone, we also investigate the presence of information in the environment.

\end{abstract}

\maketitle

\section{Introduction}
In recent years, growing attention has been given to the investigation of open quantum systems~\cite{OQSionSimulator, brownianTrappedIons, LOsimulationOpenSystems, OQSemulationSQ, InesDeVegaOpenSys, collisionBasedEvolution}, which are ubiquitous in quantum science, where many experimental systems have significant interactions with their environment \cite{Unruh1995,Biercuk2009,Monz2011,Preskill2013,Paladino2014,Malinowski2018}. %,Kawakami2016,Reina2002,Aharonov2006,Yang2016,Staudacher2013
Although their description is significantly more involved compared to isolated systems~\cite{NielsenChuang},
it allows to capture omnipresent effects as decoherence and other instances of state deterioration as well as memory effects in a unified manner.

Especially for quantum information applications, a point of particular interest is the flow of information from and to the open system, usually captured by the concept of Markovianity.
This concept is classically well-understood and identifies stochastic processes as \textit{Markovian} if they are memory-less, such that the future of the process depends only on the present state, but not on its past~\cite{WisemanMarkov,BreuerBook}.
However, a straightforward generalization of this notion to the quantum realm is not possible and thus several nonequivalent definitions exist (see, e.g., Ref.~\cite{BreuerReview}).
Among the various possibilities, the criterion formulated by Breuer, Laine and Piilo (the so-called \textit{BLP criterion}) for non-Markovianity~\cite{nonMarkovianMeasure, nonMarkovianMeasure2, review_nonMarkovianity_breuer} used in this work stands out insofar as it gives a direct operational meaning to information in open systems as the degree of distinguishability of systems states with optimal measurements.

In the field of open systems research, the focus naturally lies on the dynamics of the open system, which is analyzed in terms of a quantum master equation or effective dynamical maps~\cite{BreuerBook}.
Also the information flow and memory effects are usually expressed in terms of information present in, lost from or returned to the system. However, questions as for example where information lost from the system actually flows to, i.e., whether it actually is transferred from the system to the environment or becomes stored in correlations between the two, are usually not in focus~\cite{BreuerReview}.
A few works address this topic and at least consider system environment correlations in more detail~\cite{Maziero2012,Pernice2012,Reina2014,Roszak2015,Chen2018f,Strzalka2020}, by for example showing that it is possible to deduce from observations on the system, whether these correlations are classical or quantum. %,Salamon2017,Roszak2018,Roszak2019,Rzepkowski2020,Pernice2011,,Costa2016b,Rossatto2011,

\begin{figure}[h!]
    \centering
	\includegraphics[width=0.47\textwidth]{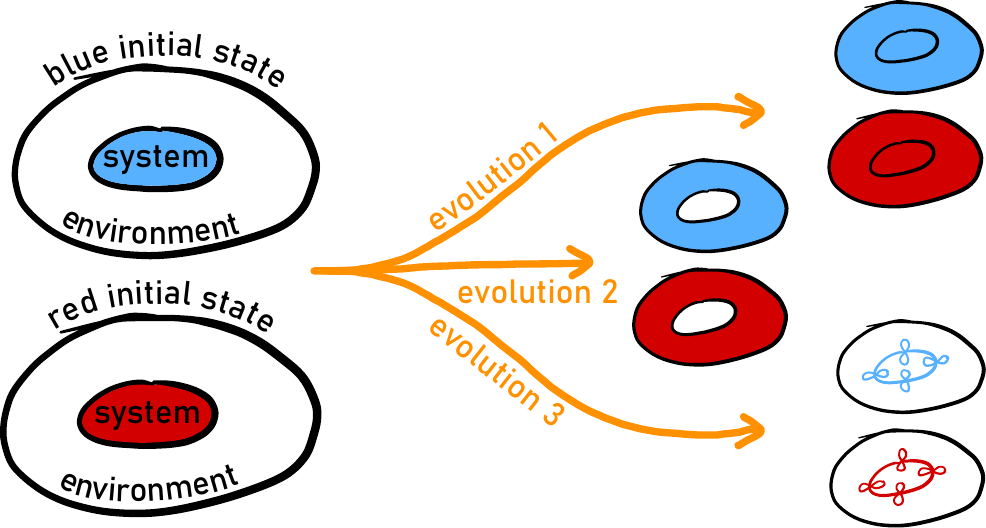}
	\caption{
	Concept of information and information flow between system and environment.
	Here, \textit{information} is defined as the distinguishability between two initial states of a quantum system (depicted as colors blue and red).
	Initially, this information is in the system only.
	After an evolution of type 1, the information from the system is duplicated into the environment such that measurements on system or environment alone are sufficient to distinguish case blue from case red.
	In contrast, after evolution 2 the information flows fully into the environment and measurements on the system alone do not enable to distinguish between the two versions.
	Finally, evolution 3 depicts a completely different case, where the information becomes stored exclusively in the correlations between system and environment, such that only a joint measurement of both allows to recover the relevant information.
	}
	\label{fig_concept}
\end{figure}

In our approach we extend the scope of investigation by explicitly asking where the information lost from the system flows to, in particular to what extent it reaches the environment, as this can have fundamental implications on memory effects and possibilities to retrieve information in application scenarios, e.g., when parts of a system and hence carriers of information are lost.
%where, e.g., a particular system might be lost and replaced by a new one, which in turn can recover the information from the environment via a non-Markovian process.
This shift in perspective allows us to analyze a taxonomy of several classes of information transfer emerging naturally, some of which are illustrated in Fig.~\ref{fig_concept}.

We investigate different types of information flow in an experimental quantum simulation using integrated photonic waveguide arrays~\cite{integratedPhotQRW, laserWrittenCircuits}.
The simulation is implemented by identifying the polarization as the quantum system and the path as the environment.
This allows to effectively realize an open system through the interaction between these two degrees of freedom by harnessing polarization-dependent coupling.
Employing laser-written structures in silica provides a high degree of control over both the path and the polarization degrees of freedom of photons, as well as the scalability necessary to match the complexity of more realistic systems.

The regular structure of our integrated circuit allows to analytically calculate the full time evolution of system and environment states, thus expanding the usual perspective focused solely on the effective evolution of the open system.
We clearly observe a non-Markovian evolution of the system qubit and investigate the flow of information to the environment.
By extending the concept of the BLP criterion we distinguish different manners in which the information is encoded in our system.
Furthermore, we propose a different type of array with a decisively different type of information flow, where all information is removed from the system and transferred to the environment.

\begin{figure*}
    %\begin{picture}(500,400)
    %\put(22,160){\includegraphics[scale=0.17]{setup8.pdf}}
    %\end{picture}
    \includegraphics[width=\textwidth]{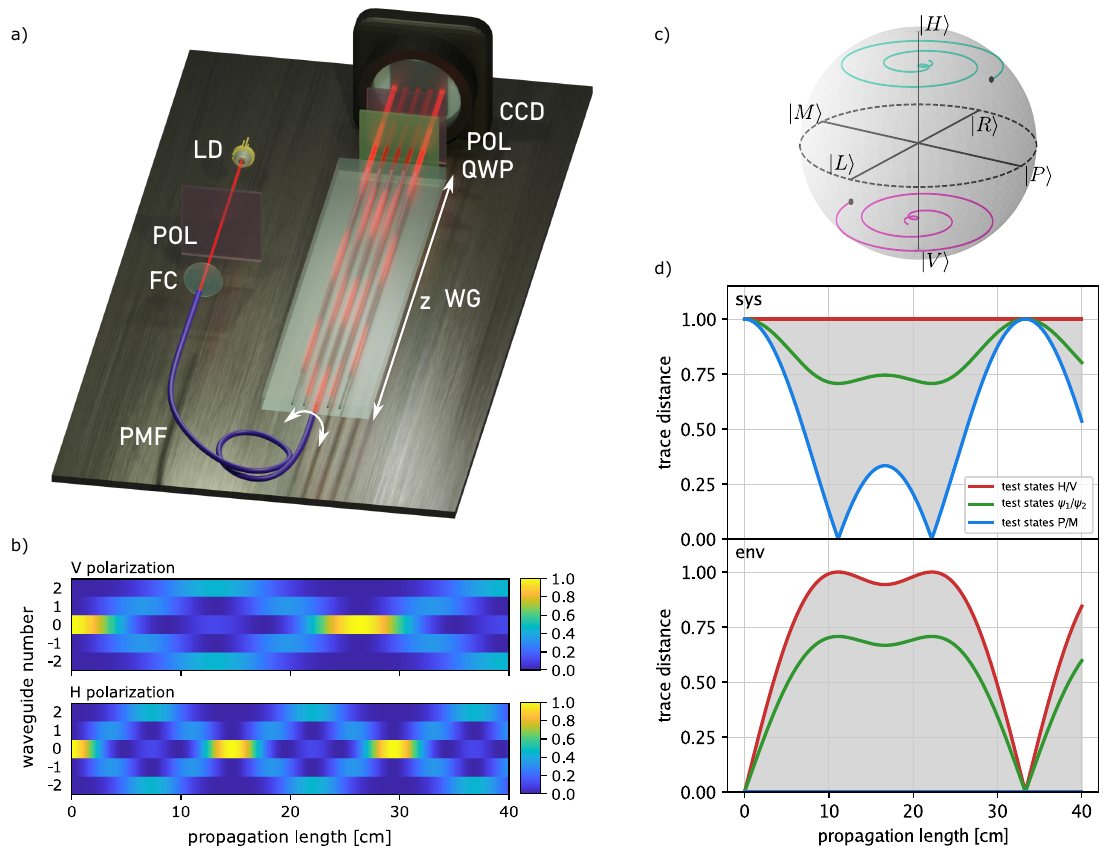}
	\caption{Experimental setup and simulations.
	a) Experimental setup consisting of a linearly polarized $780\,\rm{nm}$ laser (LD) coupled to the chip via a PM fiber.
	At the output of the waveguide chip (WG), polarization state tomography is performed using a $\lambda/4$ waveplate (QWP), a polarizer (POL) and a CCD camera.
	Waveguide chips with different lengths $z$ were used to measure at various propagation lengths.
	b) Transverse probability distributions of photons in the waveguide array as function of the propagation length for the input states $\ket{\mathrm{H}}$ and $\ket{\mathrm{V}}$.
	c) Simulated evolution of the polarization states for $\ket{\psi_1}$ (upper turquoise curve) and $\ket{\psi_2}$ (lower purple curve) inside the Bloch sphere.
	d) Simulations of polarization and path trace distance behavior for different pairs of orthogonal test states $\ket{\mathrm{H}}$ and $\ket{\mathrm{V}}$ (red curves), $\ket{\mathrm{P}}$ and $\ket{\mathrm{M}}$ (blue curves) as well as $\ket{\psi_1}$ and $\ket{\psi_2}$ (green curves). Here and in the remaining plots, the gray area signifies the possible range of trace distances under variations of pairs of orthogonal test states.
	}
	\label{fig_expSetupAndSimulations}
\end{figure*}

\section{Tracing Information}

Following the BLP criterion, we here define information as the ability to distinguish between two quantum states.
Any choice of two states $\rho_1$ and $\rho_2$ corresponds to a certain amount of distinguishing information encoded in the parameters characterizing the two states.
This distinguishability is quantified via the trace distance $D$ between the two states, defined as
\begin{align}
    D(\rho_1,\rho_2) = \frac{1}{2}\lVert \rho_1 - \rho_2 \rVert = \frac{1}{2}\tr{\sqrt{(\rho_1 - \rho_2)^2}}.
    \label{eq_TD}
\end{align}

To trace the flow of information, we consider two initially perfectly distinguishable pure test states of the system, $\rho_1$ and $\rho_2$.
For these two test states, a certain unitary interaction with the environment results in changed reduced states of system $\rho^S_i=\mathrm{tr}_{ E}{(\rho^{S\land E}_i)}$ and environment $\rho^E_i=\mathrm{tr}_{S}{(\rho^{S\land E}_i)}$, where $\rho^{S\land E}_i$ is the joint state of system and environment after the evolution in the two cases with $i = 1,2$.
After the interaction, even via optimal measurements on the system it might become harder or even impossible to decide in which of the two test states the system initially has been.
This corresponds to an information loss from the \textit{system}, quantified by the trace distance $D_S = D(\rho^S_1,\rho^S_2)$.
Conversely, after the interaction, it might become possible to infer the initial test state via optimal measurements on the \textit{environment}, indicating information transfer to it, quantified via the trace distance $D_E = D(\rho^E_1,\rho^E_2)$.

To analyze how a particular interaction transfers information, we choose different pairs of test states. 
After the interaction the pairs might be distinguishable to varying degree by measurements on the system or the environment or even allow distinguishability only from joint measurements.
This allows to characterize the interaction by its different transfer behavior of information in dependence of the choice of information encoding in test states.
This analysis also includes the original application of the BLP criterion to witness non-Markovian information backflow, in the case where, for a time-evolving interaction, $D_S$ increases with time.

\section{Experiment}
\subsection{Waveguide Arrays as Open Quantum Systems}

In order to implement the quantum simulation of an open quantum system we use polarized photons propagating in an array of evanescently coupled waveguides.
The polarization and the path degrees of freedom form a composite system equivalent to an open quantum system (polarization) coupled to an environment (path).
A sketch of the experimental setup is shown in Fig.~\ref{fig_expSetupAndSimulations}a.
The array, laser-written in fused silica ($\text{SiO}_2$)~\cite{szameit_discrete_2010}, is composed of five waveguides with a separation of $33\mu\text{m}$ lying in a plane parallel to the surface of the chip.
The waveguides have an elliptical cross section introducing birefringence into the system as well as a polarization-dependent evanescent coupling between the waveguides.

We consider only single photon processes and thus the Hamiltonian $H$ acting on the Hilbert space of system and environment $\mathcal{H}_S \otimes \mathcal{H}_E$ which best describes our scenario~\cite{integratedPhotQRW} can be written as
\begin{align}
H &= \sum_{\pi = \mathrm{H},\mathrm{\mathrm{V}}} P_\pi \otimes \Big( \sum^M_{m=1}  \beta^\pi_m P_m + \sum_{j=m\pm 1} \kappa^\pi_{j,m} \kb{m}{j} \Big),
\label{eq_hamiltonian}
\end{align}
with $M=5$ and projectors $P_\pi$ and $P_m$ on polarization $\pi$ and path $m$. $\beta^\pi_m$ and $\kappa^\pi_{jm}$ represent the propagation constant for waveguide $m$ and the coupling constant from wave\-guides $j$ to $m$ for polarization $\pi$, respectively.

Since the Hamiltonian is a pure dephasing Hamiltonian \cite{Roszak2018}, i.e, it is diagonal along the system basis of horizontal ($\mathrm{H}$) and vertical ($\mathrm{V}$) polarization, the resulting evolution
\begin{align}
U(t) &= e^{i H t} = \kb{\mathrm{H}}{\mathrm{H}} \otimes U^\mathrm{H}(t) + \kb{\mathrm{V}}{\mathrm{V}} \otimes U^\mathrm{V}(t)
\label{eq_unitary}
\end{align}
is also diagonal with effective environment unitaries $U^\pi$ given by
\begin{align}
U^\pi(t) = \exp \left[i t \Big(\sum^M_{m=1}  \beta^\pi_m P_m + \sum_{j=m\pm 1} \kappa^\pi_{j,m} \kb{m}{j} \Big) \right].
\end{align}
Thus, for our system, $\mathrm{H}$ and $\mathrm{V}$ polarizations are eigenstates of the evolution $U$ such that the total evolution can be separated into two polarization conserving parts corresponding to these states.
We numerically simulated the two intensity distributions of the two polarizations, as shown in Fig.~\ref{fig_expSetupAndSimulations}b, based on the simplest possible non-trivial model for our system where all propagation and coupling constants are the same across the waveguides with $\beta^\pi_j = \beta^\pi$ and $\kappa^\pi_{j,j\pm1} = \kappa^\pi$ for all $j$.
For this and all following simulations as well as in the experimental realization we always start with an initial path state $\rho_I$ given by coupling all light into the central waveguide.
Thus, whenever the photons are initially in a superposition of $\mathrm{H}$ and $\mathrm{V}$ polarizations the evolution according to $H$ couples system and environment.

\subsection{Simulation of Information Flow}
\begin{figure}
    \centering
	\includegraphics[width=0.47\textwidth]{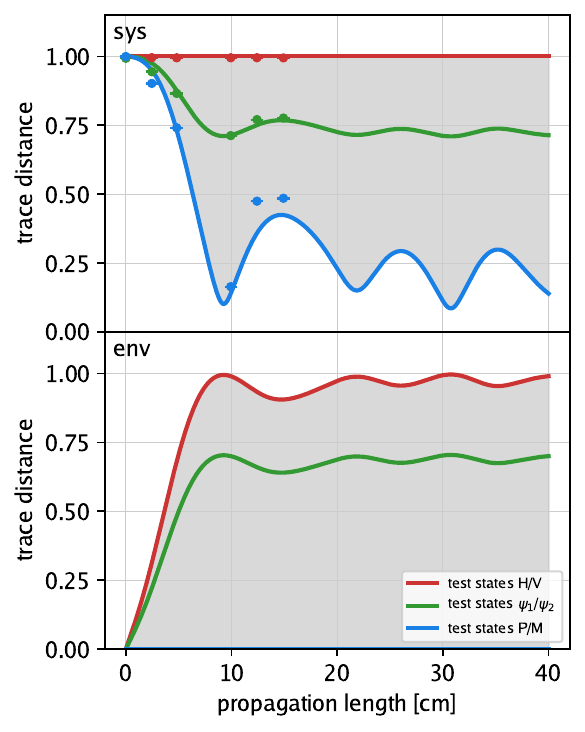}
	\caption{
	Experimental data and fit of theoretical model.
	The first datapoint of each curve at length zero stems from measurements in free space (without the array) and the remaining results are obtained using arrays of five different lengths.
	The statistical errorbars lie inside the dots.
	The solid lines represent simulations based on the theoretical model including varying propagation constants.
	}
	\label{fig_expData}
\end{figure}

Employing the trace distance Eq.~(\ref{eq_TD}) between pairs of states, it is instructive to first determine the full range of information flow\jdam{\sout{s}} between system and environment by analyzing paradigmatic cases.
Given the structure of our system, it becomes clear that no information loss is to be expected for the eigenstates $\ket{\mathrm{H}}$ and $\ket{\mathrm{V}}$ as pair of test states $\rho_1$ and $\rho_2$.
On the other hand, for a pair such as $\ket{\mathrm{P}} = (\ket{\mathrm{H}} + \ket{\mathrm{V}})/\sqrt{2}$ and $\ket{\mathrm{M}} = (\ket{\mathrm{H}} - \ket{\mathrm{V}})/\sqrt{2}$ the trace distance is expected to be most sensitive since equal weights are given to the two eigenstates of the evolution.
Additionally, we consider an intermediate case with the pair $\ket{\psi_1} = (\ket{\mathrm{H}} + \ket{\mathrm{P}}) \, / \, \sqrt{2+\sqrt{2}}$ and $\ket{\psi_2} = (\ket{\mathrm{V}} - \ket{\mathrm{M}}) \, / \, \sqrt{2+\sqrt{2}}$.

We start with a numerical simulation of the expected information flow based on the simple model described above, shown in Fig.~\ref{fig_expSetupAndSimulations}d.
Initially, we see maximum distinguishability in the system, which in the course of the evolutions drops indicating a loss of the information from the system.
For the pair $\ket{\mathrm{P}}$ and $\ket{\mathrm{M}}$, the information is completely lost.
Subsequently, however, the trace distance increases again, which witnesses non-Markovianity.
The appearance of non-Markovianity is a consequence of the form of the Hamiltonian and does not hinge on the finite boundaries of our system.
While with five waveguides we observe distinct state recurrence for $\mathrm{H}$ and $\mathrm{V}$ polarizations (Fig.~\ref{fig_expSetupAndSimulations}b) influencing the timing and strength of the information backflow, also with an infinite number of waveguides non-Markovian behavior occurs, although without any recurrence effects.

The particular structure of the information loss is furthermore illustrated for the case of $\ket{\psi_1}$ and $\ket{\psi_2}$ via the evolution of the system states in the Bloch sphere in Fig.~\ref{fig_expSetupAndSimulations}c.
It corresponds to a dephasing process in the $\{\mathrm{H},\mathrm{V}\}$-basis together with a change of the relative phase between the H and V components.
These two effects combine into the spiralling evolution of the states, where the Bloch vectors remain pointing in opposite directions.

\subsection{Measurement Results}
We perform the experiment with the same pairs of states and observe the evolution of the information in the system.
In order to achieve the time-resolved quantum state tomography of the polarization, we employ five chips of different lengths with arrays of wave\-guides written with the same writing parameters, thus leading to the same Hamiltonian.
The results are presented as data points in Fig.~\ref{fig_expData}.

Just as in our simulation the system exhibits an almost maximal information loss followed by a significant gain for the pair $\set{\ket{\mathrm{P}}$,$\ket{\mathrm{M}}}$, no significant information loss for $\set{\ket{\mathrm{H}}$,$\ket{\mathrm{V}}}$, and an intermediate behavior for $\set{\ket{\psi_1}$,$\ket{\psi_2}}$.
It is also clearly possible to certify non-Markovian behavior in the evolution.
When considering only the relative behavior of the information between the three pairs, these results are explained very well by the simple theoretical model used for Fig.~\ref{fig_expSetupAndSimulations}d. 
However, the time evolution of the information flow is noticeably different, which stems from the fact that the actual physical array exhibits variations of the parameters which are not included in the model.
A noticeable deviation from the simple model is expected to originate from small fluctuations of the laser system during the writing of the array, which result in different propagation constants for each waveguide.
Modeling these variations by allowing different parameters $\beta_k$ for the five waveguides already allows to fit an evolution which is much closer to the observed results, shown as solid lines in Fig.~\ref{fig_expData}.

\section{Information Flow in the Experiment}

To understand the different evolutions of information we observed in the experiment we provide a theoretical analysis of our system.
The state $\rho$ of the open system, i.e., the polarization qubit, is parametrized in terms of correlation tensor elements $\{ T_x,T_y,T_z\}$ as
\begin{align}
    \rho = \frac{1}{2} \left( \id + T_x \s_x  + T_y \s_y  + T_z \s_z \right),
\end{align}
with the Pauli matrices $\{\s_x,\s_y,\s_z\}$.
The time evolution of our system states (see Fig.~\ref{fig_expSetupAndSimulations}c and d as well as Fig.~\ref{fig_expData}) indicates that information encoded in $T_x$ and $T_y$ is transferred differently than information encoded in $T_z$.
%indicate that the types of information encapsulated by the parameters $T_x$ and $T_y$ behave differently than the information encoded in $T_z$.
While distinguishability along the $z$-direction of the Bloch sphere is never lost from the system, information in the $xy$-plane is maximally lost.
%Thus, it is illustrative to also express the initial pure states $\ket{\psi} \in \mathcal{H_S}$ in terms of polar angles of the Bloch sphere with
Thus, it is illustrative to choose a parametrization of the initial pure states $\ket{\psi} \in \mathcal{H_S}$ in terms of polar angles of the Bloch sphere such that
\begin{equation}
    \ket{\psi} = \cos \theta \, \ket{\mathrm{H}} + e^{i\varphi} \sin \theta \, \ket{\mathrm{V}},
\end{equation}
where $\theta \in \left[ 0,\pi/2 \right]$ and $\varphi \in \left[ 0,2\pi \right]$.
Angles and correlation tensor elements are related as $T_x = \cos \phi \sin (2\theta)$, $T_y = \sin \phi \sin (2\theta)$ and $T_z = \cos (2\theta)$.
Here, for the environment we consider more generally any pure initial state of the environment $\rho_I = P_\Phi$, with the projector $P_\Phi = \kb{\Phi}{\Phi}$ and $\ket{\Phi} \in \mathcal{H_E}$, which in general corresponds to a coupling into a coherent superposition of waveguide modes.

For an initially pure system state $\ket{\psi}$, after applying the evolution $U$ described in Eq.~(\ref{eq_unitary}) and tracing over environment, i.e., path, we obtain for the final system state $\rho^S$ at time $t$ the expression
\begin{align}
    \rho^S(t) = \begin{pmatrix}
	\cos^2 \theta & \cos \theta \, \sin \theta e^{-i\varphi} \gamma^\ast(t) \\
	\cos \theta \, \sin \theta e^{i\varphi} \gamma(t) & \sin^2 \theta
	\end{pmatrix},
	\label{eq_stateSysFinal}
\end{align}
with $\gamma(t) := \langle \Phi | (U^H(t))^\dagger U^V(t) | \Phi\rangle \in \mathbb{C}$ and $|\gamma| \in \left[ 0,1 \right]$.
Conversely, the final state of the environment $\rho^E$ can be written as
\begin{align}
    \rho^E(t) = \langle P_H \rangle \, U^H P_\Phi (U^H)^\dagger + \langle P_V \rangle \, U^V P_\Phi (U^V)^\dagger,
    \label{eq_stateEnvFinal}
\end{align}
where the expectation value is taken over the initial system state $\ket{\psi}$, with $\langle P_H \rangle = \cos^2 \theta$ and $\langle P_V \rangle = \sin^2 \theta$ and the time dependence is contained in the unitaries with $U^\pi = U^\pi(t)$.

From Eqs.~(\ref{eq_stateSysFinal}) and (\ref{eq_stateEnvFinal}) the amount of information in system and environment can be calculated in dependence on the choice of initial information encoding.
For two initial test states $\rho_1$ and $\rho_2$ parametrized by $\{T^1_x,T^1_y,T^1_z\}$ and $\{T^2_x,T^2_y,T^2_z\}$ the trace distances $D_S$ and $D_E$ between the effective system states and between the environment states after the evolution, respectively, are given as
\begin{align}
D_S(t) &= \frac{1}{2} \sqrt{\delta T^2_z + |\gamma(t)|^2 (\delta T^2_x + \delta T^2_y)}, \label{eq_distSys}\\
D_E(t) &= \frac{1}{2} \sqrt{\left(1-|\gamma(t)|^2\right) \; \delta T_z^2} \label{eq_distEnv}
\end{align}
with $\delta T_k = T^1_k - T^2_k$.
These equations hold for any unitary with a diagonal form as $U$ with arbitrary $U^H$ and $U^V$, and any pure initial environment state $\rho_I = \kb{\Phi}{\Phi}$, which together determine the parameter $\gamma$.

\section{Types of Information Flow to Environment}
\subsection{Classical Information Flow}
We can distinguish two completely different types of information flow for the information encoded either in $T_z$ or alternatively in $T_x$ and $T_y$.
As seen from Eq.~(\ref{eq_distSys}), the information encoded in $T_z$ never leaves the system and is at least partially duplicated in the environment when $\gamma > 0$.
$\varphi$-dependent information on the other hand is never stored in the environment as becomes clear from the absence of the parameters $T_x$ and $T_y$ from Eq.~(\ref{eq_distEnv}).
Consequently, this information must become stored exclusively in the correlations between system and environment, whenever it is lost from the system.
Additionally, a clear tradeoff becomes apparent.
The more information encoded in $T_z$ is transferred to the environment, the more is necessarily lost from the complementary parameters $T_x$ and $T_y$.
This is in accordance with well-known quantum complementarity relations and the no-cloning theorem~\cite{Heisenberg1930,Wootters1982,Jaeger1995,Englert1996,Ozawa2003,Busch2013,Knips2018}, which would be violated if significant information about one parameter could be extracted from the system while leaving the information stored in the other parameters intact.
At the same time the information encoded in $T_z$ remains in the system and can be even fully copied to the environment for $\gamma = 0$.
Thus, albeit the system becomes entangled with the environment, the information transfer itself in this case is \textit{classical} in the sense that in principle it could be transmitted by a classical channel after a quantum measurement.

\begin{figure}
	\centering
	\includegraphics[width=0.47\textwidth]{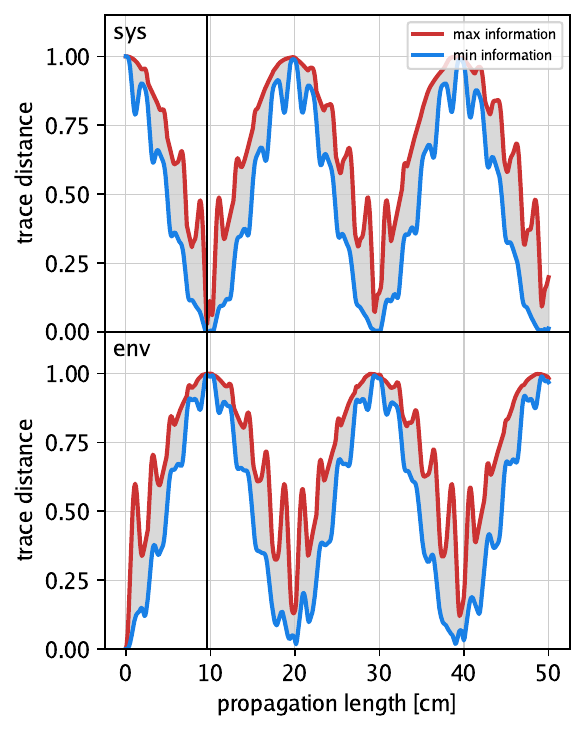}
	\caption{Evolution of information present in system and environment for specific parameters of $H^\prime$.
	The two curves in each plot are obtained via optimization (independently for system and environment) over all possible Bloch sphere directions for pairs of orthogonal test states.
	The upper curve (red) always indicates the best case trace distance, where the lower curve (blue) corresponds to the worst case.
	At propagation length $l = 9.6\,\mathrm{cm}$ (vertical marker) the unitary is (up to product unitaries) approximately equivalent to a generalized swap $U_S$, with a worst case trace distance in the environment of $0.995$.}
	\label{fig_fullTransfer}
\end{figure}

\subsection{Quantum Information Flow}

The special form of information flow in our experimental system naturally implies the question whether simple time-independent Hamiltonians such as $H$ are also suitable to implement more general types of evolutions.
In particular, a genuinely \textit{quantum} information flow occurs when a sufficiently large amount of information about several parameters is transferred into the environment such that this process could not be realized by a classical channel anymore.

The extreme of such an information flow is a full transfer of the system's quantum state into a subspace of the environment, realized by a generalized swap operation $U_\mathrm{S}$ with
\begin{align}
U_\mathrm{S} = \sum^2_{j,k=1} \kb{\phi_j}{\phi_k} \otimes \kb{\chi_k}{\chi_j} + \id \otimes P_\perp,
\end{align}
where $\{\ket{\phi_1},\ket{\phi_2}\}$ is a system basis, $\{\ket{\chi_1},\ket{\chi_2}\}$ is a pair of orthogonal environment states and $P_\perp$ is the projector into the respective orthogonal subspace of the environment.

A possible practicable solution to experimentally achieve a genuine quantum information flow to the environment, is to additionally introduce a polarization rotation into the waveguides around an axis different than $z$.
This can be achieved for example by introducing a rotation of the ellipticity into the waveguides~\cite{polRotatingWaveguides,SzameitRotation}.
The effective Hamiltonian in this case $H^\prime$ contains the Hamiltionian $H$ and an additional polarization rotation with
\begin{align}
    H^\prime = H + \sum_{m=1}^{M} (\alpha^x_m \s_x + \alpha^y_m \s_y) \otimes P_m,
\end{align}
where $\alpha^x_m$ and $\alpha^y_m$ represent the rotation rates around the $x$ and $y$-axis of the Bloch sphere for waveguide $m$ respectively.
Note that $H^\prime$ cannot be written in diagonal form for the system and thus in general also the effective evolution is not diagonal in any system basis.

By choosing suitable parameters for propagation, coupling and rotation for $H^\prime$ we are able to numerically simulate an evolution where all information about all system parameters is simultaneously transferred to the environment, as shown in Fig.~\ref{fig_fullTransfer}.
For each timestep we optimized over pairs of orthogonal test states from all possible directions of the Bloch sphere to determine the worst and best cases.
In the simulated time interval the state, and thus all information is periodically swapped between system and environment, with approximately a full swap (up to local unitaries) realized at the propagation length $l=9.6\,\mathrm{cm}$.

\section{Conclusions and outlook}

In this work, we investigated the quantum simulation of open quantum systems in optical waveguide arrays both experimentally and theoretically, which lead to a new perspective on the information flow between open systems and their environments.
By using multiple waveguide arrays with different lengths we obtained a stable window into the evolution of the system and were able to perform a time-resolving state tomography.
This allowed to observe non-Markovian evolution of the open system and to identify different types of information flow.

Conceptually, we extended the BLP measure as a general quantifier for the presence of information.
This enables to trace the information lost from our experimental system and identify our scenario as a special case of information transfer to the environment, which can be regarded as classical, in the sense that only information encoded in one type of parameter is copied to the environment.
Through further theoretical analysis we derived a tradeoff relation for our scenario, requiring information encoded in a certain parameter to be lost from the system if information stored in another parameter is gained in the environment, which we confirmed via simulations.
Conversely, we also demonstrated that the platform of optical waveguide arrays is suitable to implement a broad range of different scenarios by showing the simulation of a theoretical example with full quantum state transfer between system and environment.

Our results form the basis of a comprehensive analysis of information flow between open system and environment from a new perspective, which is sensitive to the presence of information in the environment.
In particular, it allows to investigate to which extent statements about the information present in the environment can be based on observations on the system alone.
Another possible extension of our research is the analysis of multi-photon states and their interactions with a common or with separate environments, in particular with respect to effects originating from entanglement.

\section*{Acknowledgments}
We thank Harald Weinfurter for fruitful discussions and extensive support. 
We also thank Julian Roos for constructive exchange about open quantum systems.
This work has been supported by the DFG Beethoven 2 Project No. WE2541/7-1, and by the DFG under Germany's Excellence Strategy EXC-2111 390814868.
A.S. acknowledges funding from the European Research Council (grant no. 899368 ‘EPIQUS’), the Deutsche Forschungsgemeinschaft (grants SCHE 612/6-1, SZ 276/12-1, BL 574/13-1, SZ 276/15-1, SZ 276/20-1 and SFB 1477 “Light-Matter Interactions at Interfaces”, project number 441234705), and the Alfried Krupp von Bohlen and Halbach Foundation.
J.D. acknowledges support by the international Max-Planck-Research school for Quantum Science and Technology (IMPRS-QST), and J.D.A.M. acknowledges support of the MCQST START fellowship.

\bibliography{markovBib}

\end{document}